\documentstyle[12pt]{article}
\input{psfig}
\long\def\comment#1{}
\begin{document}
\title{Three Interesting 15th and 16th Century Comet Sightings 
in Kashmiri Chronicles}

\author{Subhash Kak}

\maketitle

\begin{abstract}
This note is about three interesting
15th and 16th century sightings of comets in Kashmiri chronicles.
We provide reasons for their identification as the
1468 S1, 1531 (Halley's), and 1533 M1 comets.

\end{abstract}

\section*{Introduction}

Indian chronicles have not been properly studied for their
astronomical references. In particular, the reports of
great comet sightings can help with chronological questions
and also provide information that is useful to the astronomer.
Here we consider the medieval
Kashmiri chronicles, in particular the continuation 
of the R\={a}jatara\.{n}gi\d{n}\={\i}
by \'{S}r\={\i}vara and Pr\={a}jyabha\d{t}\d{t}a and \'{S}uka, for
their comet references.
For those who cannot read the original Sanskrit, one may consult
Jogesh Chunder Dutt's English translation (Dutt, 1898).

As background, Kashmir has a long tradition of historical writing.
The R\={a}jatara\.{n}gi\d{n}\={\i} of Kalha\d{n}a (Stein, 1900) gives an account
of kings from their mythical beginnings
until 1150.
Jonar\={a}ja, \'{S}r\={\i}vara, Pr\={a}jyabha\d{t}\d{t}a and \'{S}uka
bring the 
narrative forward to the time of the incorporation of Kashmir in the
Mughal Empire by Akbar in 1586.

The dates in the Kashmiri texts are by the Laukika (Saptar\d{s}\i)
calendar. This calendar begins with 3076 BC, the year starting
in March-April (Kak, 2000). The equation of
conversion is, therefore,

4500 Laukika = 1424-5 AD

Comets are a commonly discussed subject in the Indian literature. 
One of the longest chapters in Var\={a}hamihira's encyclopaedic
B\d{r}hat Sa\d{m}hit\={a} (505 AD) is on comets (Bhat, 1981).
A background to Indian astronomy and its relationship with Mesopotamian
and Greek astronomy is available elsewhere (Kak, 2003a, 2003b).

\section*{The Comet of 1468}

The
reign of the good king Zain-ul-abidin
(Jainoll\={a}bhad\={\i}na in Sanskrit chronicles), or
Badashah, who ruled from 1420 to 1470, is described in the books by
Jonar\={a}ja and \'{S}r\={\i}vara. Jonar\={a}ja died in 1459
and the subsequent account by \'{S}r\={\i}vara, who was the
king's minister, continues the story.

\'{S}r\={\i}vara describes his comet in the last of the
seven chapters of the book that he devoted to the king's reign. 
(He has three more books describing the successors to this king.)
This indicates that the comet made its visitation near the
end of the king's rule. The description sets the stage for the
succession struggle between the king's sons.

Here's the description of the comet in Dutt's translation:

\begin{quote}
A comet was seen at night in the north.
Its long tail was of resplendent beauty.
For a period of two months the comet was visible in the clear sky.
[Page 153]
\end{quote}

This is interspersed with the supposed calamitous effects of the comet.
For example: ``The people saw signs of a severe calamity to the
country which had hitherto been happy under good government.
[The comet] is the cause of
the destruction of men, even as excessive rain is of embankments.
The king remained anxious through fear of mischief that might happen.''

\'{S}rivara does not explicitly
mentioned the Laukika year when speaking of the comet.
But since the king died in 1470 and the succession war appears to
have taken several months, the year 1468 becomes clear. The comet appears
at about the winter harvesting, as its appearance comes at
about the same time as the entry into Kashmir of refugees from a
famine in another, unnamed country. The foreign 
refugees are likely to
have become more numerous the following spring as the
text describes.
 This is followed by a series of
events that include the 
burning of the city of Suyyapura, the death of
the queen, the death of the king's nephew in the province
of Sindhu, and the drama of the struggle
for power between his three sons over a period of several months.
This brings us to 1470, when the king dies.

In summary, the description suggests that the comet arrived in
the autumn of 1468 and it lingered for two  months.

A later passage further informs us:

\begin{quote}
An eclipse of the moon and of the sun took place within a
fortnight, as if meant to upset the king and thereby to
destroy the kingdom in which there had hitherto been no
division. [Page 154]
\end{quote}

This provides further evidence that can date the comet.
\subsection*{Confirmation of the date}

Modern tables on comets confirm that in 1468, comet S1 was first
seen on 18 September with the maximum brightness date of 2 October.
The comet was visible for 56 days, confirming the approximately stated
period of two months by \'{S}r\={\i}vara.
The magnitude of this comet is estimated to be 1-2.

Calculations of solar eclipses reveal that soon after this period
occurred the solar eclipse of 9 July 1469 centered at the latitude
of 69.3N and 134.8E with a path width of 380 km and center duration
of 4 minutes and 6 seconds. The lunar eclipse of July 24, 1469 (fifteen
days later) confirms
the other description in the text.
Although the eclipses may not have been seen by \'{S}rivara personally, 
one must
remember that the astronomy almanacs then, as now, provided dates for 
the eclipses.

\section*{The Comets of 1531 and 1533}

The comets of 1531
and 1533 are described in the book by Pr\={a}jyabha\d{t}\d{t}a and \'{S}uka.
The year and the month is mentioned. The 
first of these is sighted in the Laukika year 4607.
The text reads:

\begin{quote}
In the year 4607, K\={a}ca Cakrapati, intending to fight
with the M\={a}rgapatis, moved his army and a {\it comet appeared
in the west.} [Page 369]
\end{quote}

The Laukika 4607 is March-April 1531 to March-April 1532.
But since the Kashmiri winters are harsh, one must assume that this
movement would have occurred sometime before winter. That makes
it the year 1531.

The Laukika year 4608 (1532) has continuing war and the sacking
of the towns by foreign invaders from the northwest. Next,

\begin{quote}
In the year 9, in the month of Jye\d{s}\d{t}ha, the Mughals returned 
to their country, taking with them by force the wealth of the people, and by 
treaty the daughter of the king. In this way calamity befell the sinful
people of the Sat\={\i}sara country [Kashmir], and a {\it comet was seen
continuously  in the sky on the east and on the west.}..
Stars fell from the sky on the fields where the full harvest
of rice was ripening, and the comet became again visible. [Page 373]
\end{quote}
This second comet foreshadows a terrible famine, caused, no doubt,
by rains that destroyed the crops. This second comet, therefore,
must have been seen in the rains in July and later.

The comet 1531  was the
Halley's comet that had a maximum magnitude of 1 and was visible starting
August 5 for a period of 34 days is clearly
the one mentioned first.

The next comet appears to have been 1533 M1, with maximum magnitude 
of 0, that became visible on 
June 27, 1533 for a period of 83 days. We see how its arrival is
coincident with the ripening rice in the fields.
One may imagine that incessant rains beginning mid July, the
skies remained overcast and the rice crop was ruined.
When the rain stopped, the comet was seen again, for it remained
visible until late September.

\section*{Discussion}

We now know that another bright comet was observed in 1532, but
it does not find a mention in the Chronicle. 
This was the comet R1 of maximum magnitude -1 that became visible on
September 2 for a period of 120 days.
But this omission is understandable in light of the perilous
situation in Kashmir that year.
According to the Chronicle,
in the month of Agrah\={a}ya\d{n}a in autumn, the K\={a}sk\={a}ra
king Saida Khana sent a general with
twelve
thousand cavalry to sack Kashmir. 
The reputation for barbarity of the K\={a}sk\={a}rians (from the mountains
between east Afghanistan and Chitral) being worse than that
of the Turks, the Kashmiris simply fled their homes. 
The K\={a}sk\={a}rians attacked the twin capital cities. 
According to the Chronicle,
``Hundreds of thousands of low houses were burnt, and the brilliant city
became like the ground for burning the dead, fearful to look at its
charred wood.
Where will kings get two such capitals in which millions had been
spent in lime, wood, brick and painting?'' [Page 370-1]

The Kashmiri army regrouped in the south and the battles went
on, in one area, for three months. Ultimately, after many futile
stands the Kashmiris submitted to the invaders.

Having fled their homes, in constant danger for their lives, 
attention was not given to the arrival of the new comet in the
sky.

\paragraph{Acknowledgement}
I would like to thank Professor Narahari Achar of University of Memphis
for discussion of the eclipse dates.

\section*{Bibliography}
\begin{description}

\item
M.R. Bhat, 1981. {\it Var\={a}hamihira's B\d{r}hat Sa\d{m}hit\={a}}.
Motilal Banarsidass, Delhi.

\item
J.C. Dutt, 1898. {\it The R\={a}jatara\d{n}gi\d{n}\={\i} of Jonar\={a}ja}.
Gian Publishing House, Delhi, Reprinted 1986.

\item 
S. Kak, 2000. ``Birth and early development of Indian astronomy,''
In {\it Astronomy Across Cultures: The History of 
Non-Western Astronomy.} H. Selin (ed.). Kluwer Academic,
Boston, 2000, pp. 303-340; arXiv: physics/0101063 

\item
S. Kak, 2003a. ``Babylonian and Indian astronomy: early connections.''
arXiv: physics/0301078

\item
S. Kak, 2003b. ``Greek and Indian cosmology: review of early history.''
arXiv: physics/0303001

\item
M.A. Stein, 1900. {\it Kalha\d{n}a's R\={a}jatara\d{n}gi\d{n}\={\i}}.
Motilal Banarsidass, Delhi, Reprinted 1979.

\end{description}

\end{document}